\documentclass[preprint,amsmath,amssymb,amsfonts]{revtex4-1}

\usepackage{graphicx}
\usepackage{subcaption}
\usepackage{dcolumn}
\usepackage{bm}
\usepackage{color}

\usepackage[normalem]{ulem}  

\begin{document}

\preprint{APS/123-QED}

\newcommand{\diff}{\mathrm{d}} 
\newcommand{\divergence}{\mathrm{div}\,}  
\newcommand{\grad}{\mathrm{grad}\,}  
\newcommand{\rot}{\mathrm{rot}\,}  

\newcommand{\ad}{a^\dagger}
\newcommand{\ct}{\cos\theta}
\newcommand{\st}{\sin\theta}
\newcommand{\cct}{\cos^2\theta}
\newcommand{\sst}{\sin^2\theta}
\newcommand{\G}{\Gamma}

\renewcommand{\labelitemii}{$\circ$}


\title{Anomalous behavior of the energy gap in the one-dimensional quantum $XY$ model}

\author{Manaka Okuyama$^1$}
\author{Yuuki Yamanaka$^1$}
\author{Hidetoshi Nishimori$^1$}
\author{Marek M. Rams$^2$}%
\affiliation{%
$^1$Department of Physics, Tokyo Institute of Technology, Oh-okayama, Meguro-ku, Tokyo 152-8551, Japan
}
\affiliation{%
$^2$Institute of Physics, Jagiellonian University, \L{}ojasiewicza 11, 30-348 Krak\'ow, Poland
}
\date{\today}

\begin{abstract}

We re-examine the well-studied one dimensional spin-1/2 $XY$ model to reveal its nontrivial energy spectrum, in particular the energy gap between the ground state and the first excited state. In the case of the isotropic $XY$ model -- the $XX$ model -- the gap behaves very irregularly as a function of the system size at a second order transition point. This is in stark contrast to the usual power-law decay of the gap and is reminiscent of the similar behavior at the first order phase transition in the infinite-range quantum $XY$ model. The gap also shows nontrivial oscillatory behavior for the phase transitions in the anisotropic model in the incommensurate phase. We observe a close relation between this anomalous behavior of the gap and the correlation functions. These results, those for the isotropic case in particular, are important from the viewpoint of quantum annealing where the efficiency of computation is strongly affected by the size dependence of the energy gap.

\end{abstract}

\pacs{Valid PACS appear here}
\maketitle

\section{Introduction} 

The  ground state of a quantum many-body system changes drastically as parameters in the Hamiltonian are driven across a quantum transition point.
This phenomenon is preceded in a finite-size system by closing of the energy gap between the ground and first excited states as a function of the system size growing toward the thermodynamic limit.
The rate of the gap closing is an important measure characterizing a quantum phase transition.

According to the finite-size scaling theory, when applied to a second order phase transition
with a divergent correlation length, physical quantities generally behave
polynomially as a function of the system size \cite{nishimori-ortiz,santos,osterloh}.
In contrast, the gap is expected to close exponentially fast at the first order
quantum phase transition:
the two ground states at opposite sides of the first order transition point have
significantly different properties and consequently their
overlap in a finite-size system is very -- typically exponentially --  small.
The overlap of the two states determines the energy gap since the overlap
corresponds to the off-diagonal elements of the effective two-level Hamiltonian
describing the system around the transition point and the gap is
directly related to the magnitude of the off-diagonal elements.
It is therefore expected that the order of quantum phase transition
in the thermodynamic limit is generally in one-to-one correspondence
with the rate of the gap closing toward the thermodynamic limit,
polynomially for the second order transition and exponentially for the first order transition \cite{privman,hatano}.

Studies of the energy gap are also important from the viewpoint of quantum annealing \cite{kadowaki,kadowaki2,finilla,morita,das,santoro,bapst},
in which parameters of the system are controlled in time and driven across a quantum critical point.
The rate of the gap closing directly affects the efficiency of quantum annealing.
An exponential closing of the gap implies an exponentially long computation time whereas a polynomial gap leads to a polynomial time \cite{farhi,seki,seoane,bapst2,jorg,young1,young2}.

Although the above-mentioned correspondence between the order of quantum phase
transition and the rate of the gap closing generally holds true in most systems, an
interesting counterexample has been found by Cabrera and Jullien
\cite{cabrera,cabrera2} who showed that the first order quantum phase
transition in the one-dimensional transverse-field Ising model with
antiperiodic boundary condition is accompanied by a polynomial closing of the energy gap. 
See also Ref. \cite{laumann} for essentially the same result.
Furthermore, another very anomalous example has been given for the infinite-range
quantum $XY$ model, where the energy gap behaves in many different ways at first order
quantum phase transitions \cite{tsuda}.
It has been shown there that many types of the gap closing: polynomial, exponential,
and even factorial closing coexist along a critical line in the phase diagram.
This example suggests that we should be very careful while relating
the type of quantum phase transition to the rate of the gap closing.

In the present paper, we first analyze the one-dimensional quantum isotropic $XY$ model with $s=1/2$, which exhibits he second order phase transition, 
and show that this model is another example of the anomalous gap behavior.
Although the properties of one-dimensional quantum spin systems
have been studied from a number of different perspectives
\cite{sachdev,Henkel,lieb,katsura,pfeuty,mccoy,kurmann,hoeger,kenzelmann,antonella,Damski,CV2010,CNPV2014,CPV2015},
our focus is on the anomalous system-size dependence of the gap,
at variance from most of the previous studies.
We conclude that the energy gap behaves quite peculiarly (highly oscillatory) as a function of the system size,
in a very similar way to the case of the infinite-range quantum $XY$ model \cite{tsuda}.
This observation should be taken seriously from the perspective of efficiency of quantum annealing.

Next, we analyze the anisotropic $XY$ model and show  that the energy gap also behaves anomalously as a function of the system size in some regions of the parameters.
Finally, we observe that a direct relation holds between the energy gap and the correlation function in the ground state. For instance, the anomalous oscillatory behavior of the gap precisely coincides with the oscillations of the correlation functions. 

The organization of the paper is as follows. In Sec. I\hspace{-.1em}I, we define the model, diagonalize the Hamiltonian and obtain the energy gap.
In Sec. I\hspace{-.1em}I\hspace{-.1em}I, we analyze the size dependence of the energy gap using both analytical and numerical methods. In Sec. I\hspace{-.1em}V, we show that a relation is established between the energy gap and the correlation function in the systems. Finally, our conclusion is given in Sec. V.

\section{The Hamiltonian and its spectrum}	
We study the one-dimensional quantum $XY$ model with $s=1/2$ in transverse and longitudinal fields, $\Gamma$ and $h$, respectively, and periodic boundary condition,
\begin{eqnarray}
H&=&-\frac{1}{2} \sum_{i=1}^{N}[(1+\gamma)\sigma_i^x\sigma_{i+1} ^x+(1-\gamma)\sigma_i^y\sigma_{i+1} ^y]-\Gamma \sum_{i=1}^{N} \sigma_i^z-h\sum_{i=1}^{N} \sigma_i^x \label{eq:hamiltonian1},
\end{eqnarray}
where $\gamma$ controls the anisotropy, and $\sigma_i^x, \sigma_i^y$ and $\sigma_i^z$ are the Pauli matrices acting at site $i$. For the moment, we assume that the system size $N$ is even.
Without loss of generality, we assume that $\Gamma \ge 0$ and $\gamma \ge0$.

We consider quantum annealing protocol where the longitudinal filed $h$ is driven in time for a given, fixed value of the transverse field $\Gamma$.
For $\gamma>0$, there is a first order quantum transition at $h=0$ between the phases with $\langle \sigma_j^x \rangle>0$ and $\langle \sigma_j^x \rangle<0$  (for $h<0$ and $h>0$, respectively), as long as $0\leq \Gamma < 1$, see Fig. 1 \cite{lieb,katsura,pfeuty,mccoy,kurmann,hoeger,kenzelmann,antonella}.
\begin{figure}[!htb]
\begin{center}
\includegraphics[width=0.4\hsize]{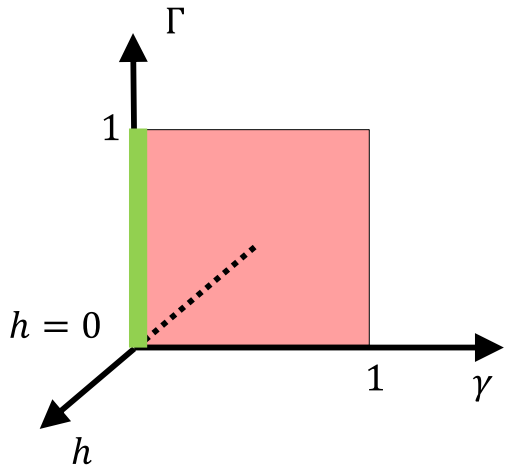}
\end{center}
\caption{The phase diagram of the XY model in Eq.\ (\ref{eq:hamiltonian1}), which undergoes a first order phase transition across the surface marked in pink 
between the phases with positive and negative expectation values of $\sigma_j^x$. Along the isotropic line of $\gamma=0$ marked in green, for $0\le \Gamma <1$, the transition  is of the second order.}
\label{fig:phasediagram}
\end{figure}
This expectation value has a jump at $h=0$, which is the reason why we call it the first order transition.
In contrast, for $\gamma=0$, a second order transition takes place, as the
expectation value $\langle \sigma_j^x \rangle$ continuously changes across 0 at $h=0$ \cite{mccoy2}.

Our aim is to clarify the system-size dependence of the energy gap at $h=0$.
Notice that the longitudinal field in Eq.\ (\ref{eq:hamiltonian1}), $-h\sum_j\sigma_j^x$, induces transitions between states with even and odd values of
the $z$-component of the total spin $S_{\rm tot}^z=\frac{1}{2}\sum_j\sigma_j^z$, while other terms in the Hamiltonian conserve the parity. 
For our protocol, it is therefore relevant to study the energy gap between the ground
and first excited states, where one has an odd and the other has even value of $S_{\rm tot}^z$.

The XY model in Eq.\ (\ref{eq:hamiltonian1}), for $h = 0$, can be diagonalized following the standard procedure \cite{lieb,katsura,pfeuty,mccoy,kurmann,hoeger,kenzelmann,antonella}. 
Since the energy gap depends on the subtle finite-size effects in the energy spectrum, we repeat the derivation in some detail.
After the Jordan-Wigner transformation, $\sigma^z_i= 2 c_i^\dagger c_i-1  $, $\sigma^x_i - i \sigma^y_i = 2 c_i \prod_{j<i}(-\sigma^z_j)$, where $c_i$ are fermionic annihilation operators, the Hamiltonian is expressed as
\begin{eqnarray}
H=&-& \sum_{i=1}^{N-1} \left[(c_i^{\dagger}c_{i+1}-c_ic_{i+1}^{\dagger})+\gamma(c_i^{\dagger}c_{i+1}^{\dagger}-c_ic_{i+1}) \right]
\nonumber \\ 
&+& (-1)^{N_c} \left[ (c_N^{\dagger}c_{1}-c_N c_{1}^{\dagger})+\gamma(c_N^{\dagger} c_{1}^{\dagger}-c_N c_{1}) \right]
\nonumber \\ 
&-&2 \Gamma \sum_{i=1}^{N} (c_i^{\dagger} c_i -\frac{1}{2}) \label{eq:hamiltonian2} ,
\end{eqnarray}
where we will call $N_c \equiv \sum_i c_i^{\dagger} c_i$ sign operator.
After the Fourier transformation, 
\begin{eqnarray}
c_j= \frac{1}{\sqrt{N}} \sum_{k}e^{ikj} a_k ,
\end{eqnarray}
the Hamiltonian in Eq.\ (\ref{eq:hamiltonian2}) is transformed as
\begin{eqnarray}
H&=& \sum_{k} H_k ,
\\
H{_k}&=& - \left[ 2a_k^{\dagger} a_k (\cos k +\Gamma) + \gamma ( e^{ik}a_k^{\dagger} a_{-k}^{\dagger} - e^{-ik}a_k a_{-k}) \right] +  \Gamma \label{k4}  , 
\end{eqnarray}
where the values of $k$ depend on the boundary condition: $k = k_1$ for periodic $c_{N+j} = c_j$ ($N_c$ odd) and $k = k_2$ for antiperiodic $c_{N+j} = -c_j$ ($N_c$ even), with
\begin{eqnarray}
k_1 &=&0 , \pm \frac{2}{N} \pi ,\cdots ,  \pm \frac{N-2}{N} \pi , \pi \label{k_1} ,
\\
k_2 &=& \pm \frac{1}{N} \pi , \pm \frac{3}{N} \pi , \cdots ,  \pm \frac{N-1}{N} \pi \label{k_2} .
\end{eqnarray}
For $\gamma=0$, the transformed Hamiltonian $H_k$ is already diagonal. The part of the Hamiltonian for a given absolute value of momentum reads
\begin{eqnarray}
H_k + H_{-k} = &-2&
\left[ 
\begin{array}{ccc}
a_k^{\dagger} && a_{-k}^{} 
\end{array} 
\right]
\left[ 
\begin{array}{ccc}
\cos k +\Gamma && i\gamma \sin k \\
-i\gamma \sin k && -\cos k - \Gamma
\end{array} 
\right]
\left[ 
\begin{array}{ccc}
a_k\\
a_{-k}^{\dagger}
\end{array} 
\right] -2 \cos k.
\end{eqnarray}
We diagonalize it using the Bogoliubov transformation
\begin{eqnarray}
\left[
\begin{array}{ccc}
d_k  \\
d_{-k}^{\dagger} \\
\end{array} 
\right]
&=& \left[
\begin{array}{ccc}
\cos \frac{\theta_k}{2} && i \sin \frac{\theta_k}{2} \\
i \sin \frac{\theta_k}{2} && \cos \frac{\theta_k}{2} \\
\end{array} 
\right]
\left[
\begin{array}{ccc}
a_k  \\
a_{-k}^{\dagger} \\
\end{array} 
\right],
\\
\nonumber \\
\cos \theta_k &=& \left(\cos k +\Gamma\right) / \epsilon(k), \label{eq:costh}\\ 
\sin \theta_k &=& \gamma \sin k / \epsilon(k) \label{eq:sinth},\\
\label{epsilonk}
\epsilon (k) &=&  \sqrt{ (\cos k+\Gamma)^2 + (\gamma \sin k)^2} \label{epsilon} .
\end{eqnarray}
Finally, the diagonalized Hamiltonian in each parity sector reads

(i) For $N_c$ odd 
\begin{eqnarray}
H^{\mathrm{odd}} &=& -2 (\Gamma+1) d_0^{\dagger} d_0 - 2(\Gamma-1) d_{\pi}^{\dagger} d_{\pi} + 2\Gamma 
\nonumber \\
&&- 2\sum_{k_3}  ( d_{k_3}^{\dagger} d_{k_3} +d_{-k_3}^{\dagger} d_{-k_3}  -1 ) \epsilon (k_3) , \\
k_3 &=& \frac{2}{N} \pi , \frac{4}{N} \pi , \cdots, \frac{N-2}{N} \pi .
\end{eqnarray}
The ground-state energy is
\begin{eqnarray}
E_0^{\mathrm{odd}} &=& -2 - 2\sum_{k_3} \epsilon(k_3)  = -2 - \sum_{k_1} \epsilon(k_3) +\epsilon(0)+\epsilon(\pi) \nonumber
\\
&=&
\begin{cases}
-\sum_{k_1} \epsilon(k_1) & (\Gamma \le 1)
\\
2(\Gamma-1) -\sum_{k_1} \epsilon(k_1) & (\Gamma>1)  
\end{cases} \label{eq:E_0^{odd}},
\end{eqnarray}
where we have to be careful to pick the state with correct fermionic parity.

(ii) For $N_c$ even 
\begin{eqnarray}
H^{\mathrm{even}} &=& -2\sum_{k_4}  ( d_{k_4}^{\dagger} d_{k_4} +d_{-k_4}^{\dagger} d_{-k_4}  -1 ) \epsilon (k_4), \\
k_4 &=& \frac{1}{N} \pi ,  \frac{3}{N} \pi , \cdots, \frac{N-1}{N} \pi .
\end{eqnarray}
The ground-state energy is
\begin{eqnarray}
E_0^{\mathrm{even}} = - 2\sum_{k_4} \epsilon(k_4) = -\sum_{k_2} \epsilon(k_2) .
\end{eqnarray}
The true ground-state energy is given by one of these two possibilities, $E_0^{\mathrm{odd}}$ or $E_0^{\mathrm{even}}$, while the energy of the first excited state is the other one. Therefore the energy gap is equal
\begin{eqnarray}
\Delta(N,\Gamma,\gamma) \equiv  \left| E_0^{\mathrm{odd}}-E_0^{\mathrm{even}} \right| &=\begin{cases}
\displaystyle{\left| \sum_{k_2} \epsilon(k_2)-\sum_{k_1} \epsilon(k_1) \right| }& (\Gamma \le 1)
\\
\displaystyle{2(\Gamma-1)  + \sum_{k_2} \epsilon(k_2) -\sum_{k_1} \epsilon(k_1)} & (\Gamma>1)  
\end{cases} \label{eq:E_0^{odd}}   \label{eq:Delta2}
\end{eqnarray}

When the system size $N$ is odd, we diagonalize the Hamiltonian in the same way and obtain the energy gap as,
\begin{eqnarray}
\Delta(N,\Gamma,\gamma) =
\begin{cases}
\displaystyle{ \left| \sum_{k_5} \epsilon(k_5)-\sum_{k_6} \epsilon(k_6)  \right| } & (\Gamma \le 1)
\\
\displaystyle{2(\Gamma-1) +\sum_{k_5} \epsilon(k_5) -\sum_{k_6} \epsilon(k_6)   }& (\Gamma>1)  
\end{cases}  \label{eq:Deltaodd}
\end{eqnarray}
where
\begin{eqnarray} 
k_5 &=&0 , \pm \frac{2}{N} \pi ,\cdots ,  \pm \frac{N-3}{N} \pi , \pm \frac{N-1}{N} \pi \label{k_5} ,
\\
k_6 &=&\pi, \pm \frac{1}{N} \pi , \pm \frac{3}{N} \pi , \cdots ,  \pm \frac{N-2}{N} \pi \label{k_6},
\end{eqnarray}
are wave numbers in the subspaces with odd and even parities, respectively. 

\section{Energy gap as a function of the system size} 
In this section, we derive the asymptotic form of the energy gap and show that it behaves in an irregular way as a function of the system size. 

\subsection{Isotropic case $(\gamma = 0)$}
We start with the isotropic case of $\gamma = 0$, for which the Hamiltonian is already diagonal in Eq.\ (\ref{k4}),
\begin{eqnarray}
H&=& \sum_{k}   \left[ -2a_k^{\dagger} a_k (\cos k +\Gamma) +  \Gamma \right] \label{eq:H} ,
\end{eqnarray}
where the values of $k$ depend on the boundary condition as in Eqs.\ (\ref{k_1}) and (\ref{k_2}). We find it convenient not to use the general expression derived in Eq.\ \eqref{eq:Delta2} here, but we start with the above formula instead. For simplicity, we only consider even $N$ here.

We carefully identify the ground state in preparation to evaluate the energy gap.
An apparent candidate for the ground state is the state where all modes with wave numbers satisfying
\begin{align}
\cos k > -\Gamma \label{concon},
\end{align}
are occupied.

\begin{figure}[!htbp]
\begin{center}
\includegraphics[height=55mm]{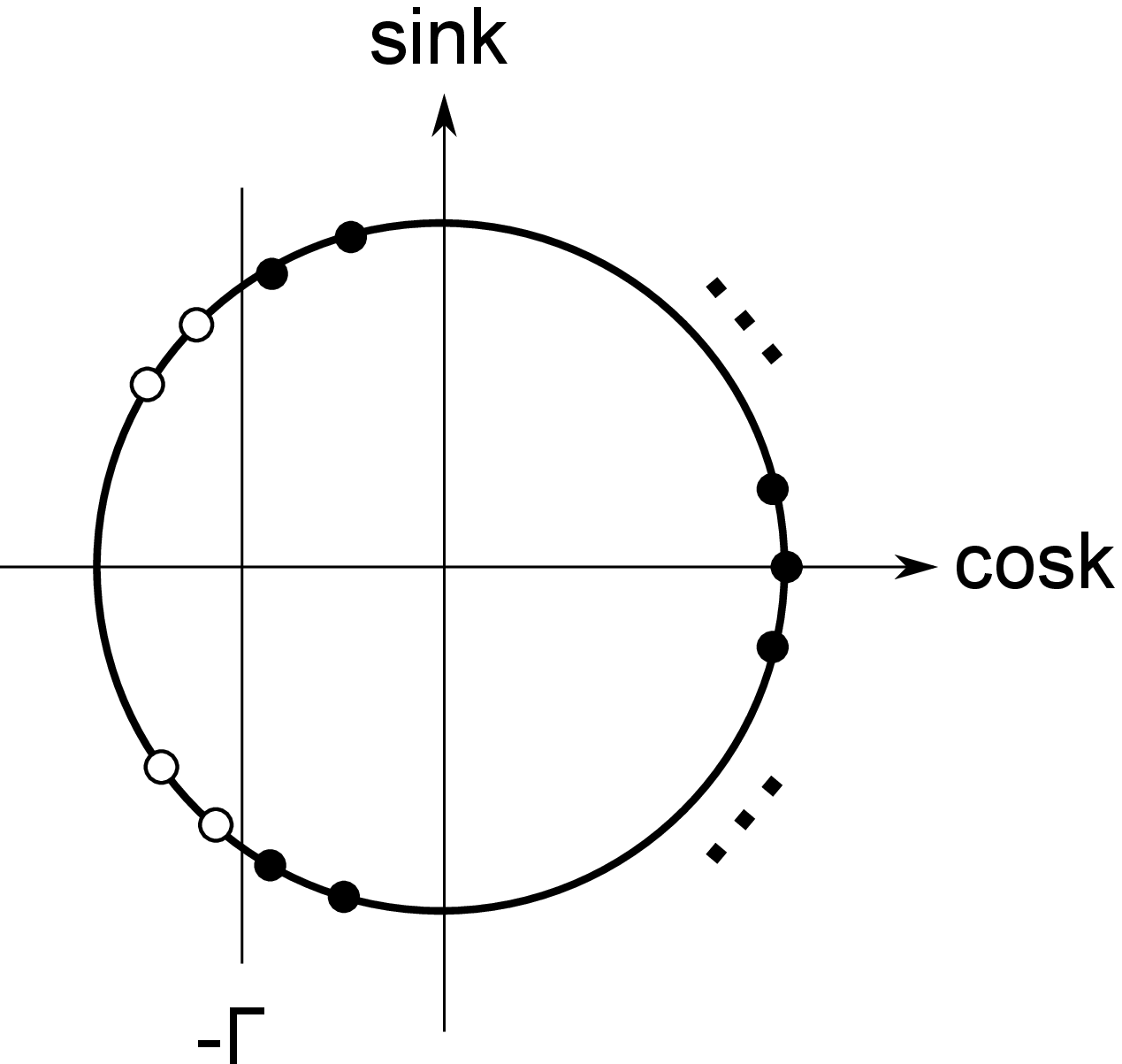} \hspace{1.5cm}
\includegraphics[height=55mm]{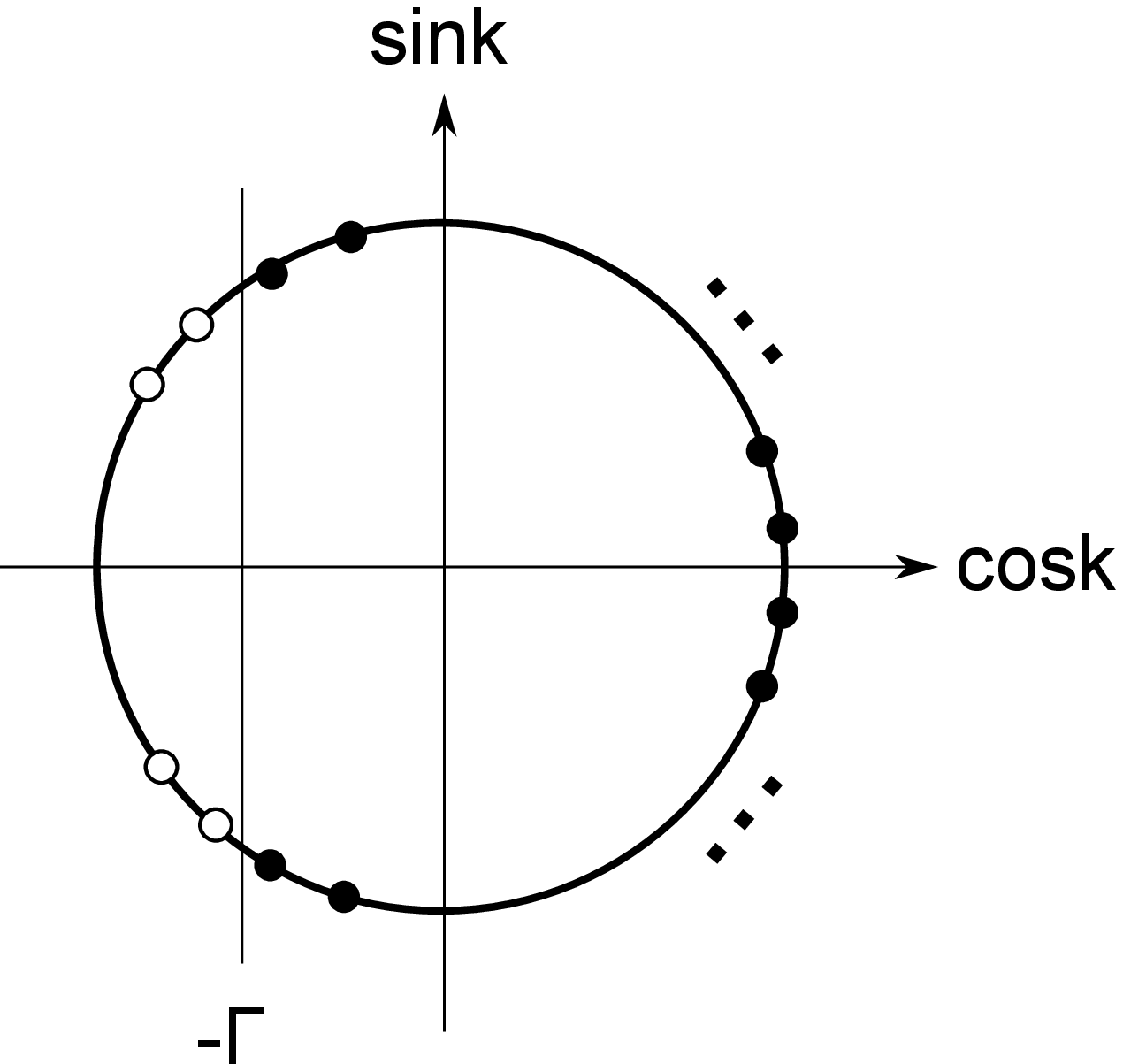}
\end{center}
\caption{Left panel: Wave numbers for the presumed ground state in the case of $k_1$ on the
complex-$k$ plane. The black points mark the occupied modes, and the white ones are for the unoccupied ones. The number of occupied modes is odd.
Right panel: Likewise, the number of occupied states for $k_2$ is even.}
\label{fig:k1}
\end{figure}

We check that those states have the number of fermions $N_c$ matching the respective parity sectors.
Indeed, as should be clear from Fig. \ref{fig:k1}, the series of wave numbers $k_1$ in
Eq.\ (\ref{k_1}) is consistent with odd $N_c$ and $k_2$ is consistent with even $N_c$. 
Notice that the parity of $N_c$ coincides with the parity of $S_{\rm tot}^z$ up
to a constant $N/2$,
\begin{equation}
S_{\rm tot}^z=\frac{1}{2}\sum_{j=1}^N \sigma_j^z
=\sum_{j=1}^N \Big(c^{\dagger}_jc_j -\frac{1}{2}\Big) =N_c-\frac{N}{2},
\end{equation}
which is related to the statement that we consider the gap between states with even and odd values of $S_{\rm tot}^z$.

\subsubsection{Energy gap}

The lowest energy in each parity sector is given by Eq.\ (\ref{eq:H}), with $a_k^\dagger a_k =1$ for $k$ satisfying Eq.\ (\ref{concon}) and $a_k^\dagger a_k=0$ otherwise. 
The energy gap reads
\begin{align}
\Delta (N,\Gamma,0)= \left| \sum_{|k_1| \le \frac{2 n }{N} \pi} \left[ -2\left(\cos k + \Gamma \right) \right] 
- \sum_{|k_2| \le \frac{2 m -1}{N}\pi} \left[ -2\left(\cos k + \Gamma \right) \right] \right| , \label{gap}
\end{align} 
where $\frac{2 n }{N}\pi$ and $\frac{2 m -1}{N}\pi$ are, respectively, the largest $k_1$ and $k_2$ satisfying Eq.\ (\ref{concon}). 
Then,
\begin{align}
\sum_{|k_1| \le \frac{2 n }{N} \pi} \cos k = 1 + 2\sum_{j=1}^n \cos \frac{2j}{N} \pi = \frac{\displaystyle\sin \frac{(2n+1)\pi}{ N}} {\displaystyle\sin\frac{\pi}{N}}; ~ ~ ~ ~
\sum_{|k_2| \le \frac{2 m -1}{N}\pi} \cos k = \frac{\displaystyle\sin \frac{2m\pi}{N}}{\displaystyle\sin \frac{\pi}{N}}.
\label{k1sum}
\end{align}
Next, we find $n$ and $m$, as
\begin{align}
n= \left\lfloor \frac{Nx}{2\pi} \right\rfloor
,~~
m= \left\lfloor \frac{Nx}{2\pi} +\frac{1}{2} \right\rfloor
,~~
x = \arccos (-\G).
\label{nm}
\end{align}
Dividing the quantities in the floor symbol above into integer and fractional parts,
\begin{align}
\frac{Nx}{2\pi} &= \left\lfloor \frac{Nx}{2\pi} \right\rfloor + \delta_1 
= n + \delta_1 ~~~(0\leq\delta_1<1),  \label{Nx2pi}\\
\frac{Nx}{2\pi} + \frac{1}{2} &= \left\lfloor \frac{Nx}{2\pi} 
+ \frac{1}{2} \right\rfloor + \delta_2 = m + \delta_2 ~~~(0\leq \delta_2<1) .
\end{align}
we find that for $0\leq\delta_1<\frac{1}{2}$, $\delta_2=\delta_1+\frac{1}{2}$, and the energy gap is 
\begin{align}
\Delta(N,\Gamma,0)=
2 \left|
\frac{\displaystyle\G\cos\left(\frac{\pi}{2N}(1-4\delta_1)\right)+\sqrt{1-\G^2}
\sin\left(\frac{\pi}{2N}\left( 1-4\delta_1 \right)\right)}
{\displaystyle\cos \frac{\pi}{2N}}-\G \right|,
\label{kekka1} 
\end{align}
while for $\frac{1}{2} \leq\delta_1<1$, $\delta_2=\delta_1-\frac{1}{2}$ and the gap reads
\begin{align}
\Delta(N,\Gamma,0) = 2 \left| \frac{\displaystyle\G\cos\left(\frac{\pi}
{2N}(3-4\delta_1)\right)+\sqrt{1-\G^2}
\sin\left(\frac{\pi}{2N}\left(3-4\delta_1 \right)\right)}
{\displaystyle\cos \frac{\pi}{2N}}-\G \right|.  \label{kekka2}
\end{align}
In both cases we can expand the above expressions in the powers of $1/N$, obtaining the asymptotic formula
\begin{align}
\Delta(N,\Gamma,0) \approx \frac{\pi \sqrt{1-\Gamma^2}}{N}    \left| 2 ~{\rm mod} \left( \frac{N \phi}{\pi} ,1 \right) -1 \right| +\mathcal{O}(N^{-2}),
\label{eq:isotropic_asymptotic}
\end{align}
where the angle $\phi$ is defined as 
\begin{eqnarray}
\phi = \arccos(\Gamma).
\end{eqnarray}
The standard, polynomial decay of the gap is modified here by $\left(2 ~{\rm mod} \left( \frac{N \phi}{\pi} ,1 \right) -1\right)$, which is a piecewise linear, continuos function of $\phi N$, oscillating between $-1$ and $+1$. 

In the Appendix A we present another derivation of the above formula using more general approach, suitable for anisotropic case as well.

\subsubsection{Anomaly in the energy gap}
\begin{figure}[t]
\begin{minipage}[]{.5\linewidth}
\centering
\includegraphics[width=1.0\hsize]{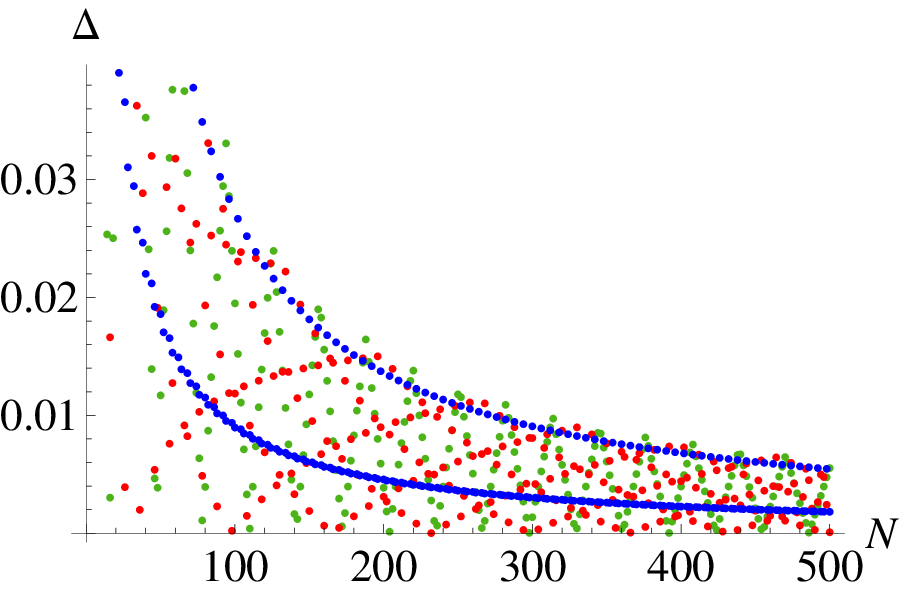}
\end{minipage}%
\begin{minipage}[]{.5\linewidth}
\centering
\includegraphics[width=1.0\hsize]{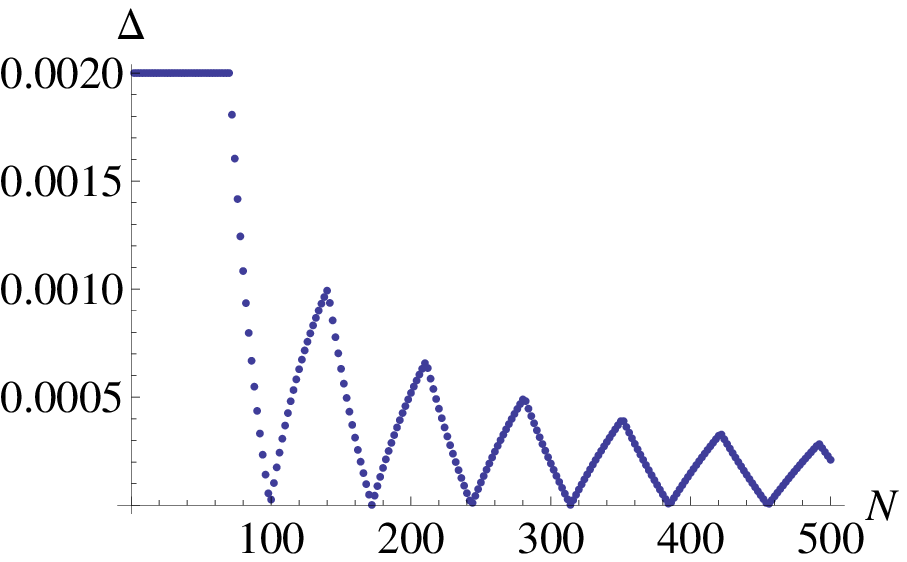}
\end{minipage}%
\caption{System-size dependence of the energy gap for the isotropic XY model in Eq.\ \eqref{eq:hamiltonian1} with $\gamma=0$ and $h=0$. Left panel: Green, red and blue dots show values of the energy gap for $\Gamma=0.1$, $\Gamma=0.3$ and $\Gamma=0.5$, respectively. Right panel: The energy gap for $\Gamma=0.999$.}
\label{fig:initial}
\end{figure}

We plot typical examples of the energy gap as a function of the system size in Fig. \ref{fig:initial}, which have been calculated using the exact formulas in Eqs.\ (\ref{kekka1}) and (\ref{kekka2}).

\begin{figure}[b]
\begin{center}
\includegraphics[width=0.5\hsize]{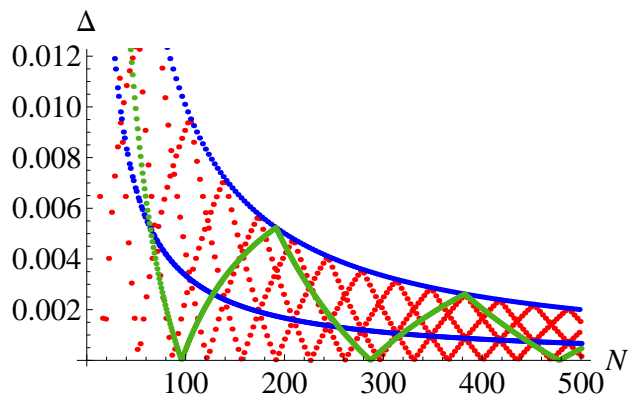}
\end{center}
\caption{System-size dependence of the energy gap for the infinite-range XY model in Eq.\ \eqref{hamiltonian_IR} for the longitudinal field $h$=0 and various, fixed values of the transverse field $\Gamma$. Red, blue and green dots show the data for $\Gamma=\pi/10$, $\Gamma=1/3$ and $\Gamma=1-\pi/300$, respectively.}
\label{fig:IR}
\end{figure}

Most interesting, these figures look qualitatively very similar to those for the infinite-range quantum $XY$ model \cite{tsuda},
\begin{align}
H_\infty = -\frac{1}{4N} \sum_{i,j=1}^{N}
\left(\sigma^x_i \sigma^x_{j} + \sigma^y_i \sigma^y_{j} \right)
- \frac{\Gamma}{2} \sum_{i=1}^{N} \sigma^z_i
- \frac{h}{2} \sum_{i=1}^{N} \sigma^x_i   \label{hamiltonian_IR}.
\end{align}
For convenience, we reproduce some of them in Fig. \ref{fig:IR}.

It is worth remaining that, in the infinite-range XY model, the gap can be tuned
to close polynomially, exponentially, or factorially by using an appropriate series of system sizes \cite{tsuda}.
Comparison of Figs. \ref{fig:initial} and \ref{fig:IR} strongly suggests that the same may be true for the one-dimensional model as well.
We discuss it in some detail below.

\subsubsection{Sensitivity of the energy gap to the change of magnetic field $\Gamma$}

There are several important points to notice in these results.
Firstly, a slight change in $\Gamma$ can lead to drastically different behavior.
In order to illustrate this, it is useful to reduce the exact formulas for the gap in Eqs.\ (\ref{kekka1}) and (\ref{kekka2}) for some specific values of $\Gamma$.

For example, for $\Gamma =\frac{1}{2}$, $x=\arccos(-\Gamma)=2\pi /3$, and hence $Nx/2\pi =N/3$.
Then, for $N=3l~(l\in\mathbb{N})$, $\delta_1=0$ according to Eq.\ (\ref{Nx2pi}) and the gap takes a simple form
\begin{equation}
\Delta(N=3l, \Gamma=1/2, 	0) =\sqrt{3}\, \tan \frac{\pi}{2N}. \label{simple delta}
\end{equation}
On the other hand, when $N=3l+1$, $\delta_1=1/3$ and for $N=3l+2$, $\delta_1=2/3$ and the gap becomes respectively
\begin{eqnarray}
\Delta(N=3l+1, \Gamma=1/2, 0) &=& 2 \left| \frac{\displaystyle\frac{1}{2}\cos\frac{\pi}
{6N}-\frac{\sqrt{3}}{2}\sin\frac{\pi}
{6N}}{\displaystyle\cos \frac{\pi}{2N}}-\frac{1}{2} \right| \label{3l+1}, \\
\Delta(N=3l+2, \Gamma=1/2, 0) &=& 2 \left| \frac{\displaystyle\frac{1}{2}\cos\frac{\pi}
{6N}+\frac{\sqrt{3}}{2}\sin\frac{\pi}
{6N}}{\displaystyle\cos \frac{\pi}{2N}}-\frac{1}{2} \right|. \label{3l+2}
\end{eqnarray}

For large $N$, Eqs.\ (\ref{3l+1}-\ref{3l+2}) have the same system-size dependence since $\cos{\left(\pi/6N\right)} \gg \sin{\left(\pi/6N \right)}$.  
Thus, the gap follows essentially two separate curves as a function of the system size, given by Eqs.\ (\ref{simple delta}) and (\ref{3l+1}) (or (\ref{3l+2})), as can indeed be seen
in Fig. \ref{fig:initial} for $\Gamma=0.5$.

A similar argument applies to $x=\arccos (-\Gamma)=\pi l/j$ with $l, j\in\mathbb{N}$ satisfying $1/2\leq l/j<1$, the latter condition coming from $0\le \Gamma <1$.
In such a case
\begin{equation}
\frac{Nx}{2\pi}=\frac{lN}{2j},
\end{equation}
and $\delta_1$ assumes a fixed value for selected series of system sizes.
For instance, for $N=ji~(i\in\mathbb{N})$, $\delta_1=0$ or $\frac{1}{2}$,
and for $N=ji+1$, $\delta_1=l/(2j)$ or $1/2+l/(2j)$.
Each of the series $N=ji$ and $N=ji+1$ gives a smooth curve describing $\Delta$.
Similar argument applies to other series of the form $N=ji+{\rm const}$.
Therefore, for some rational values of $\arccos (-\Gamma)/\pi$, the gap, as a function
of the system size, follows a finite number of different curves. 
For irrational $\arccos (-\Gamma)/\pi$ this argument does not apply
and the gap behaves less regularly.

\subsubsection{Envelope of the energy gap}
\begin{figure}[!htbp]
\begin{center}
\includegraphics[width=0.5\hsize]{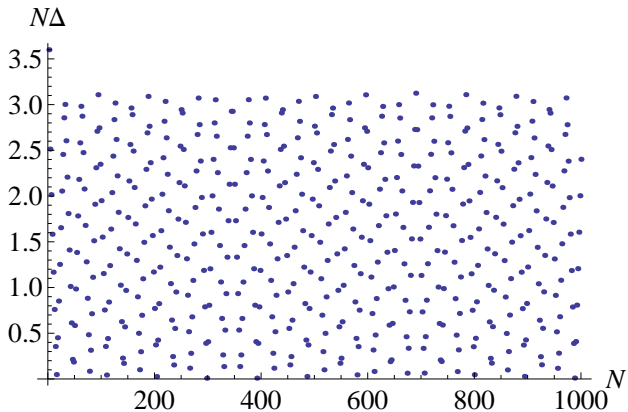}
\end{center}
\caption{System-size dependence of the gap multiplied by the size for $\gamma=0$ and $\G =0.1$.}
\label{Ndelta}
\end{figure}

Secondly, the {\it envelope} of the  gap as a function of the system size is proportional to its inverse. This is apparent when looking at the asymptotic expression in Eq.\ \eqref{eq:isotropic_asymptotic}, and we show it as well in Fig.  \ref{Ndelta} by plotting $N\Delta$ as a function of $N$ for $\Gamma =0.1$.

This power-law dependence is indeed what is expected from the second order phase transition. One can easily construct the series of the system sizes which exhibits such a behavior by  sticking to $N$'s for which the oscillating term in Eq.\ \eqref{eq:isotropic_asymptotic} is of the order of one. This would be a {\it typical} behavior as well.

\subsubsection{Possibility of rapid decay of the gap}
Note however, that we may be able to choose such a series of system sizes that the gap closes very rapidly.
The reason is that the expressions in Eqs.\ (\ref{kekka1}) and (\ref{kekka2}) oscillate between positive and negative values as illustrated in Figs. \ref{fig:initial} and in the asymptotic formula in Eq.\ \eqref{eq:isotropic_asymptotic}. This implies that the gap is crossing zero if $N$ is allowed to take continuous values. Then, it may be possible to choose appropriate values of $N$ close to those zeros which give very small, possibly exponentially small, values of the gap.
If this is correct, which we failed to prove rigorously (unlike the case of
the infinite-range quantum $XY$ model \cite{tsuda}), the rate of the gap closing
could be tuned to behave exponentially by an appropriate choice of the
series of system size, or be outright zero for some specific system size  and magnetic field $\Gamma$.

\subsubsection{Special case: independence of the system size}
Finally, Eq.\ (\ref{gap}) simplifies considerably when $\cos (\pi/N)<\Gamma$. In that a case, the summations over $k_1$ and $k_2$ in Eq.\ (\ref{gap}) run over
all allowed values of wave numbers except for $k_1=\pi$ and the gap reduces to a simple form \cite{antonella},
\begin{align}
\Delta(N, \Gamma, 0) = \left| \sum_{k_1}  \left[ -2\left(\cos k + \Gamma \right) \right] 
-2(1-\Gamma)
- \sum_{k_2} \left[ -2\left(\cos k + \Gamma \right) \right] \right|
= 2(1-\G).
\end{align} 
This last expression is independent of $N$ as can be observed in the right panel of Fig. \ref{fig:initial} in the range of $\cos (\pi/N)<\Gamma$.

\subsection{Anisotropic case $(0<\gamma<1)$}
In the anisotropic case, we derive asymptotic formulas for the energy gap and show that its behavior differs remarkably in various regions in the $(\gamma,\Gamma)$ plane. 
Most intriguingly, we observe that the energy gap is oscillating in the incommensurate ferromagnetic phase, which is neighboring the isotropic critical line discussed earlier. 
We are able to link the period of those oscillations with the oscillations of the correlation functions in those cases.

To that end, we start with Eqs.\ \eqref{eq:Delta2} and \eqref{eq:Deltaodd} and extend the procedure used in Ref. \cite{Damski} for the case of the Ising model, see Appendix A for details of the derivation. 
We summarize the results below.

\noindent In the the ferromagnetic phase for $\Gamma<1$, which is the most interesting for us, 
\begin{align}
\Delta \approx \displaystyle {\left\{ \begin{array}{ll}
\displaystyle{   \left(  \frac{-8\gamma(\Gamma^2 +\gamma^2-1-\Gamma \gamma\sqrt{\Gamma^2+\gamma^2-1})}{\pi (1-\gamma^2)} \right)^{1/2} \frac{\lambda_2^{-N}}{\sqrt{N} }   \left(1 + \mathcal{O}(N^{-1}) \right) }, & (\text{$\Gamma^2 + \gamma^2>1$), }  \\
\displaystyle{ 0},  & (\Gamma^2 + \gamma^2 =1), \\
\displaystyle{ \left| \frac{4\sqrt{2}}{\sqrt{\pi}} \left( \frac{\gamma^2(1-\Gamma^2)(1-\Gamma^2-\gamma^2)}{(1-\gamma^2)} \right)^{1/4} \frac{\alpha^N}{\sqrt{N}} \cos(\psi N+\psi_0/2)  \left(1 + \mathcal{O}(N^{-1}) \right) \right| }, & (\Gamma^2 + \gamma^2 <1).
\end{array} \right.  } \label{asymptotic form}
\end{align}
\noindent  In the paramagnetic phase for $\Gamma>1$,
\begin{equation}
\Delta \approx \displaystyle{ 2(\Gamma-1) +  \left(\frac{8\gamma \left( \Gamma^2+\gamma^2-1-\gamma \Gamma \sqrt{\Gamma^2 +\gamma^2	 -1} \right) }{\pi(1-\gamma^2)}  \right)^{1/2}  \frac{\lambda_{2}^{N}}{\sqrt{N}}  \left(1 + \mathcal{O}(N^{-1}) \right) },
\end{equation}
and, finally, on  the critical line for $\Gamma=1$,
\begin{equation}
\Delta \approx \frac{\gamma\pi}{2 N}   + \left(2\gamma- \frac{3}{2\gamma} \right) \frac{\pi^3}{48N^3} +\mathcal{O}(N^{-5}). \label{eq:gapIsing}
\end{equation}
For convenience, $\alpha$, $\lambda_2$, $\psi$ and $\psi_0$ appearing above are defined as in Ref. \cite{mccoy}
\begin{eqnarray}
\alpha &=& \sqrt{\frac{1-\gamma}{1+\gamma}}, \label{eq:alpha}
\\
\cos \psi &=& \frac{\Gamma}{(1-\gamma^2)^{\frac{1}{2}}}, \label{eq:cospsi}
\\
\psi_0 &=& \arg (\sqrt{1-\Gamma^2-\gamma^2}+i\gamma \Gamma),
\\
\lambda_2 &=& \frac{\Gamma - \left[ \Gamma^2 + \gamma^2 -1 \right]^{\frac{1}{2} } }{1-\gamma}. \label{eq:lambda2}
\end{eqnarray}
From these results, we notice that the energy gap oscillates as a function of $N$ for $\Gamma^2+\gamma^2<1$, which means that the ground state changes between subspaces with even and odd parity of $N_c$ as the system size $N$ is increasing.
On the other hand, the ground state for $\Gamma^2 + \gamma^2 >1$ always has even parity. (For $\Gamma >0$, $\gamma > 0$. For negative $\Gamma$ or $\gamma$ one can easily map the model onto the one with positive values of the parameters by suitable spin rotation. This leads to change of parity of the ground state for odd $N$ and $\Gamma<0$).

\begin{figure}[b]
\begin{center}
\includegraphics[width=0.5\hsize]{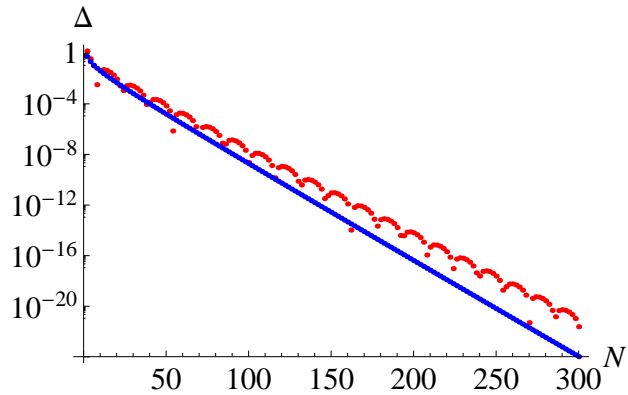}
\caption{System-size dependence of the energy gap for an anisotropic XY model in Eq.\ \eqref{eq:hamiltonian1} for $h=0$ and in the ferromagnetic phase. 
Blue and red dots show the energy gap for $(\Gamma,\gamma)=(0.85,0.95)$ and $(\Gamma,\gamma)=(0.2,0.15)$, respectively. In the second case the gap falls toward zero very rapidly at periodically observed dips.}
\label{fig:3-3}
\end{center}
\end{figure}

We plot the typical examples of the energy gap as a function of the system size in the ferromagnetic phase in Fig. \ref{fig:3-3}, which have been obtained by numerically evaluating Eq.\ (\ref{eq:Delta2}).
For $(\Gamma,\gamma)=(0.85,0.95)$, i.e. $\Gamma^2+\gamma^2>1$, the gap decays in a standard, exponential way. By contrast, for $(\Gamma,\gamma)=(0.2,0.15)$, i.e. $\Gamma^2+\gamma^2<1$, there are additional dips visible, which mark the oscillations. 

\section{Relation between energy gap and correlation functions}

We notice that the connected correlation function in the infinite system,
\begin{eqnarray}
G_{z}^R &\equiv& \langle \sigma_z^i \sigma_z^{i+R}  \rangle - \langle \sigma_z^i \rangle \langle \sigma_z^{i+R} \rangle, 
\end{eqnarray} 
has similar qualitative behavior as the gap.  We focus on the ZZ correlation function here, but this observation is independent of the specific choice of local observables. 
For convenience, we quote the asymptotic of the correlation function, following Ref. \cite{mccoy}.

\noindent In the ferromagnetic phase, $\Gamma<1$,
\begin{align}
G_{z}^R \approx \displaystyle {\left\{ 
\begin{array}{ll}
\displaystyle{ - \frac{1}{2\pi}  \lambda_2^{-2R-2}  \left(1 + \mathcal{O}(R^{-1}) \right), } &( \Gamma^2 + \gamma^2>1),  
 \\[5mm]
\displaystyle{0},  & (\Gamma^2 + \gamma^2=1 ), 
\\[5mm]
\displaystyle{  -\frac{4}{\pi} \frac{\alpha^{2R}}{  R^{2} }  \Re \left[ e^{i\psi (R+1)} \left(\frac{1-e^{2i\psi} }{1-\alpha^2 e^{-2i\psi} } \right)^{1/2}  \right]  }
\\
\displaystyle{\Re \left[ e^{i\psi (R-1)} \left(\frac{1-\alpha^2 e^{-2i\psi} }{1-e^{-2i\psi} } \right)^{1/2}  \right] \left(1 + \mathcal{O}(R^{-1}) \right), }  & (\Gamma^2 + \gamma^2 <1),
\end{array} 
\right. \label{eq:soukankansuu} }
\end{align}
where $\Re$ denotes the real part and $\alpha, \lambda_2$ and $\psi$ are defined in Eqs.\ \eqref{eq:alpha}-\eqref{eq:lambda2}.

\noindent In the paramagnetic phase, $\Gamma>1$,
\begin{equation}
G_{z}^R \approx \displaystyle{  -\frac{1}{2\pi}  \frac{\lambda_{2}^{2R}}{R^2} } \left(1 + \mathcal{O}(R^{-1}) \right).
\end{equation}
For the critical line, $\Gamma=1$,
\begin{equation}
G_{z}^R \approx \displaystyle{  -\frac{4}{\pi^2R^2} } \left(1 + \mathcal{O}(R^{-2}) \right),
\end{equation}
and finally, for the isotropic case of $\gamma=0$ and $\Gamma < 1$.
\begin{equation}
G_{z}^R = \displaystyle{  -\frac{1}{\pi^2R^2} } \sin(\psi R )^2. \label{eq:Cisotropic}
\end{equation}

Comparison of Eqs.\ \eqref{asymptotic form}-\eqref{eq:gapIsing} and \eqref{eq:soukankansuu}-\eqref{eq:Cisotropic} reveals close similarities. 
In the ferromagnetic phase, which is a first order transition from the viewpoint of our quantum annealing, the gap disappears in an exponential way on the length scale given by the correlation length in the system, which, of course, agrees with the standard qualitative prediction  \cite{sachdev}. 

Note however, that the period of oscillations of the gap as a function of the system size in the incommensurate phase ($\gamma^2 + \Gamma^2 <1$) coincides {\it precisely} with period of oscillations of the correlation function. Remarkably, the same observation holds true for the second order transition in the isotropic case. Indeed, for $\gamma = 0$, $\phi$ in Eq.\ \eqref{eq:isotropic_asymptotic} is exactly equal to $\psi$ in Eq.\ \eqref{eq:Cisotropic}. Trivially, this relation is also satisfied in other discussed cases, where neither the gap nor the correlation functions oscillate. 

For the XY model discussed in this article this relation is not accidental and can be understood since both the correlation function and the energy gap in the finite system are closely related to the Fourier coefficients of  $\cos \theta_k$  ($\sin \theta_k$) in Eqs.\ (\ref{eq:costh}-\ref{eq:sinth}) and of $\epsilon_k$ in Eq.\ \eqref{epsilonk}, respectively  (see \cite{mccoy} and the Appendix A).
Asymptotic behavior of the Fourier coefficients in both cases is determined by the poles of $\epsilon(k)$ in the complex plane, namely by $\lambda_2$. When $\lambda_2$ is real and positive there is no oscillatory behavior. On the other hand, for complex $\lambda_2$, its absolute value, $\alpha$, is giving the rate of decay and its phase, $\psi$, determines the period of oscillation.

It is an interesting question if such a relation holds true also for other models, especially since it can severally spoil the performance of the quantum annealing protocol.
While not directly relevant for us, we note, that for the systems in the thermodynamic limit there is a general and similarly looking relation linking the minima of the dispersion relation of the translationally invariant Hamiltonian (i.e. its spectrum) to the oscillations of the correlation function in the ground state \cite{Zauner2015}.

We discuss some further similarities between the gap and the correlation function in the {\it finite} system in the Appendix B. In particular, if $\Gamma^2+\gamma^2<1$ and the Fermion number is odd, a direct relation holds between the gap and the correlation function for finite size $N$ as follows,
\begin{eqnarray}
G_{z}^R(N , \Gamma , \gamma ) &\equiv& \langle \sigma_z^i \sigma_z^{i+R}  \rangle - \langle \sigma_z^i \rangle \langle \sigma_z^{i+R} \rangle ,
\\
G_{z}^N(2N , \Gamma ,\gamma ) &=& \left( \frac{1}{2N}  \frac{\partial}{\partial \Gamma} \Delta(N , \Gamma , \gamma)  \right)^2.
\end{eqnarray}
This establishes a direct nontrivial relation between the correlation function and the energy gap.

\section{Conclusion} 
In this papers, we have analytically studied the energy gap of the one-dimensional quantum $XY$ model with $s=1/2$.

We first analyzed the energy gap of the isotropic $XY$ model to
show that the energy gap at the second order transition point behaves quite anomalously as a
function of the system size toward the thermodynamic limit.
This resembles the property of the energy gap of the infinite-range quantum $XY$ model,
where the gap decreases in many different ways: polynomially, exponentially, or
factorially, depending on the choice of the series of system sizes \cite{tsuda}.
Also, our expressions of the gap, Eqs.\ (\ref{kekka1}) and (\ref{kekka2}), 
reproduce some of the results conjectured from numerical calculations in Ref. \cite{CV2010}.
In particular, their Eqs.\ (B3) and (B4) are corroborated by  the asymptotic expansion in  Eq.\ \eqref{eq:isotropic_asymptotic}. 
 
It is an interesting problem to study what aspects of these one-dimensional
and infinite-range models are the key to the anomalous size dependence of the gap.
The continuous symmetry is one of the common features of these models, but we should
carefully investigate if this is the most essential property.
From the viewpoint of implementation of quantum annealing, it is reassuring that
the energy gap can be tuned to decrease polynomially as a function of the system
size even when the lower bound closes very quickly, because the right choice of the
series of system size allows us to avoid the problematic rapid closing of the gap.
This also means that that one should be careful to interpret the results
obtained from numerical simulations or other methods for a limited series
of system sizes because different properties may emerge for different series of system sizes.

Finally, we have analyzed the energy gap of the anisotropic $XY$ model at the first order transition point.
We obtained closed asymptotic expressions for the gap and notice similarities between the energy gap and the correlation function. 
For instance, both for the anisotropic and isotropic case the oscillations of the energy gap in the finite system precisely coincides with the oscillations of the correlation function in the thermodynamic limit. It is an interesting future problem to investigate  whether or not similar relations hold in other models.

\subsection*{Acknowledgement} 
We thank Junichi Tsuda, Kazuya Kaneko, Yuya Seki, Helmut Katzgraber and Ettore Vicari for useful discussions and comments. 
The work of H.~N. was supported by the grant No. 26287086
from the Japan Society for the Promotion of Science and the work of M.M.R. by the Polish National Science Centre (NCN) grant DEC-2013/09/B/ST3/00239.

\appendix
\section{Evaluation of the energy gap} 
For the Ising model $(\gamma=1)$,  a closed analytical form of the energy gap is obtained in Ref. \cite{Damski}. 
We generalize the method to the $XY$ model.

Expanding $\epsilon(k)$ appearing in Eqs.\ \eqref{eq:Deltaodd} and \eqref{eq:Delta2} as a Fourier series $\sum_{l=0}^{\infty}a_l \cos (lk)$, we can write the energy gap as \cite{Damski}: 
\begin{eqnarray}
\Delta &=& \begin{cases}
\left| -2N(a_N +a_{3N}+a_{5N}+\cdots) \right|  & (\Gamma \le 1)
\\
2(\Gamma-1) -2N(a_N +a_{3N}+a_{5N}+\cdots) & (\Gamma>1)  
\end{cases} ,
\end{eqnarray} 
where
\begin{eqnarray}
a_l &=& \frac{1}{\pi} \int_{-\pi}^{\pi}dk \cos(kl) \sqrt{(\Gamma- \cos k)^2 +\gamma^2 \sin ^2 k}
\label{al2} .
\end{eqnarray} 

\subsection{Ferromagnetic phase, $\Gamma < 1$, with $\Gamma^2 +\gamma^2>1$}
If we introduce $z=e^{ik}$, Eq.\ (\ref{al2}) is equivalent to
\begin{eqnarray}
a_l &=& \frac{1}{\pi} \oint_{|z|=1}^{} \frac{dz}{i} z^{l-1}  \sqrt{ 1-\gamma^2} \sqrt{ \frac{(z-\lambda_1)(z-\lambda_2)(z-\lambda_1^{-1})(z-\lambda_2^{-1}) }{4z^2}  }  ,
\end{eqnarray}
with branch cuts along $(\lambda_1^{-1},\lambda_{2}^{-1}) \cup (\lambda_2,\lambda_1)$, where $\lambda_1$ and $\lambda_2$ are defined as follows,
\begin{eqnarray}
\lambda_1 &=&\frac{\Gamma + \left[ \Gamma^2 + \gamma^2 -1 \right]^{\frac{1}{2} } }{1-\gamma}
\\
\lambda_2 &=& \frac{\Gamma - \left[ \Gamma^2 + \gamma^2 -1 \right]^{\frac{1}{2} } }{1-\gamma} .
\end{eqnarray}
By deforming the integration contour around the branch cuts, we obtain
\begin{eqnarray}
a_l &=& -\frac{\sqrt{ 1-\gamma^2}}{\pi} \lambda_{2}^{-l} \int_{{\lambda_2}/{\lambda_1}}^{1}  t^{l-2}  (1-t)^{1/2}(t-\lambda_2/\lambda_1)^{1/2} (\lambda_1 \lambda_2 -t)^{1/2} (1-t/\lambda_{2}^{2})^{1/2}dt . \label{A6}
\end{eqnarray}
For large $l$, the integral is dominated by the contributions around $t=1$. We further assume that we are away from the line $\Gamma^2+\gamma^2=1$, that is $\lambda_2/\lambda_1^{}<1$, and expand the integrand around $t=1$ taking the leading order in $t-1$. We arriving at
\begin{eqnarray}
a_l &\approx&  -\frac{\sqrt{ (1-\gamma^2)}}{\pi} \lambda_{2}^{-l} (1-\lambda_2/\lambda_1)^{1/2} (\lambda_1 \lambda_2 -1)^{1/2} (1-1/\lambda_{2}^{2})^{1/2} \int_{{\lambda_2}/{\lambda_1}}^{1}  t^{l-2}  (1-t)^{1/2} dt 
\nonumber \\
&=& -\frac{\lambda_{2}^{-l}}{\pi}  \left(-\frac{8\gamma \left( \Gamma^2+\gamma^2-1-\gamma \Gamma \sqrt{\Gamma^2 +\gamma^2	 -1} \right) }{(1-\gamma^2)}  \right)^{1/2} \int_{{\lambda_1}/{\lambda_2}}^{1}  t^{l-2}  (1-t)^{1/2} dt
  .
\end{eqnarray}
We can evaluate the above integral as follows,
\begin{eqnarray}
\int_{{\lambda_1}/{\lambda_2}}^{1}  t^{l-2}  (1-t)^{1/2} dt \approx \int_{0}^{1} t^{l-2}  (1-t)^{1/2} dt \approx \frac{\sqrt{\pi}}{2l\sqrt{l}}.
\end{eqnarray}
Therefore,
\begin{eqnarray}
a_N &\approx& -\frac{\lambda_{2}^{-N}}{ \sqrt{\pi}}  \left(-\frac{8\gamma \left( \Gamma^2+\gamma^2-1-\gamma \Gamma \sqrt{\Gamma^2 +\gamma^2	 -1} \right) }{(1-\gamma^2)}  \right)^{1/2}  \frac{1}{2N\sqrt{N}}
\end{eqnarray}
and for $N$ large enough,
\begin{eqnarray}
a_N \gg a_{3N} \gg a_{5N}\cdots .
\end{eqnarray}

\noindent Using the above results, we obtain the asymptotic behavior of the energy gap as
\begin{eqnarray}
\Delta &\approx& -2N a_N  \approx \left(-\frac{8\gamma \left( \Gamma^2+\gamma^2-1-\gamma \Gamma \sqrt{\Gamma^2 +\gamma^2	 -1} \right) }{\pi(1-\gamma^2)}  \right)^{1/2}  \frac{\lambda_{2}^{-N}}{\sqrt{N}}.
\end{eqnarray}

\subsection{Critical line, $\Gamma=1$}
Taking the limit of $\Gamma \rightarrow 1$ in Eq.\ (\ref{A6}) gives
\begin{eqnarray}
a_l &=& -\frac{\sqrt{ 1-\gamma^2}}{\pi}  \int_{\frac{1-\gamma}{1+\gamma}  }^{1}  t^{l-2}  (1-t)\left(t-\frac{1-\gamma}{1+\gamma} \right)^{1/2} \left(\frac{1+\gamma}{1-\gamma}  -t \right)^{1/2} dt .
\end{eqnarray}
For large $l$, the integral is dominated by the contributions around $t=1$. We expand the integrand around $t=1$ and take the leading order in $t-1$, arriving at
\begin{eqnarray}
a_l  &=& -\frac{\sqrt{ 1-\gamma^2}}{\pi}  \int_{0}^{1}  t^{l-2}  \left[\frac{2\gamma(1-t)}{ \sqrt{1-\gamma^2}}-\frac{\gamma(1-t)^2}{\sqrt{1-\gamma^2}} -\frac{(1-t)^3}{4\gamma\sqrt{1-\gamma^2}} -\frac{(1-t)^4}{8\gamma \sqrt{1-\gamma^2}}
\right. \nonumber \\ && \left. 
+\mathcal{O}((1-t)^5) \right] dt  +\left(\frac{1-\gamma}{1+\gamma} \right)^l\mathcal{O}(l^{-1})
\nonumber \\
&\approx& -\frac{2\gamma}{\pi}   \frac{1}{l^2-l} +\frac{2\gamma}{\pi (l^3-l)}  +\frac{3}{2\pi\gamma} \frac{1}{l^4+2l^3-l^2-2l} +\frac{3}{\pi\gamma}  \frac{1}{l^5+5l^4+5l^3-5l^2-6l} +\mathcal{O}(l^{-6})
\nonumber\\
&=&-\frac{2\gamma}{\pi}   \frac{1}{l^2}  +\left( \frac{3}{2\pi\gamma} -\frac{2\gamma}{\pi}\right) \frac{1}{l^4}  +\mathcal{O}(l^{-6}) .
\end{eqnarray}
Therefore, we obtain
\begin{eqnarray}
\Delta &\approx& -2N \sum_{s=1,3,5,\dots} \left[-\frac{2\gamma}{\pi}   \frac{1}{s^2N^2}  +\left( \frac{3}{2\pi\gamma} -\frac{2\gamma}{\pi}\right) \frac{1}{s^4N^4} +\mathcal{O}(N^{-6})\right]
\nonumber\\	
&=&  \frac{\gamma\pi}{2 N}   + \left(2\gamma- \frac{3}{2\gamma} \right) \frac{\pi^3}{48N^3} +\mathcal{O}(N^{-5}) .
\end{eqnarray}

\subsection{Paramagnetic phase, $\Gamma>1$}
If we introduce $z=e^{ik}$, Eq.\ (\ref{al2}) is equivalent to
\begin{eqnarray}
a_l &=& \frac{1}{\pi} \oint_{|z|=1}^{} \frac{dz}{i} z^{l-1}  \sqrt{ 1-\gamma^2} \sqrt{ \frac{(z-\lambda_1)(z-\lambda_2)(z-\lambda_1^{-1})(z-\lambda_2^{-1}) }{4z^2}  }  ,
\end{eqnarray}
with branch cuts along $(\lambda_1^{-1},\lambda_{2}^{}) \cup (\lambda_2^{-1},\lambda_1)$.

\noindent By deforming the integration contour, we are lead to
\begin{eqnarray}
a_l &=& -\frac{\sqrt{ 1-\gamma^2}}{\pi} \lambda_{2}^{l} \int_{{1}/{\lambda_1 \lambda_2}}^{1}  t^{l-2}  (1-t)^{1/2}(t-1/\lambda_1 \lambda_2)^{1/2} (\lambda_1 /\lambda_2 -t)^{1/2} (1-\lambda_{2}^{2} t)^{1/2}dt .
\end{eqnarray}
For large $l$, the integral is dominated by the contributions around $t=1$. We expand the integrand around $t=1$ and take the leading order in $t-1$, arriving at
\begin{eqnarray}
a_l &\approx&  -\frac{\sqrt{ 1-\gamma^2}}{\pi} \lambda_{2}^{l} (1-1/\lambda_1\lambda_2)^{1/2} (\lambda_1/ \lambda_2 -1)^{1/2} (1-\lambda_{2}^{2})^{1/2} \int_{1/{\lambda_1}{\lambda_2}}^{1}  t^{l-2}  (1-t)^{1/2} dt 
\nonumber \\
&=& -\frac{\lambda_{2}^{l}}{\pi}  \left(\frac{8\gamma \left( \Gamma^2+\gamma^2-1-\gamma \Gamma \sqrt{\Gamma^2 +\gamma^2	 -1} \right) }{1-\gamma^2}  \right)^{1/2}  \int_{1/{\lambda_1}{\lambda_2}}^{1}  t^{l-2}  (1-t)^{1/2} dt 
\nonumber \\
 &\approx& -\frac{\lambda_{2}^{l}}{ \sqrt{\pi}}  \left(\frac{8\gamma \left( \Gamma^2+\gamma^2-1-\gamma \Gamma \sqrt{\Gamma^2 +\gamma^2	 -1} \right) }{(1-\gamma^2)}  \right)^{1/2}  \frac{1}{2l\sqrt{l}}.
\end{eqnarray}
Similarly as for the ferromagnetic case discussed above, for $N$ large enough the asymptotic of the gap is determined by $a_N$.
From the above results, we obtain the asymptotic behavior of the energy gap as
\begin{eqnarray}
\Delta &\approx& 2(\Gamma-1) +  \left(\frac{8\gamma \left( \Gamma^2+\gamma^2-1-\gamma \Gamma \sqrt{\Gamma^2 +\gamma^2	 -1} \right) }{\pi(1-\gamma^2)}  \right)^{1/2}  \frac{\lambda_{2}^{N}}{\sqrt{N}}.
\end{eqnarray}

\subsection{Incommensurate ferromagnetic phase, $\Gamma^2 +\gamma^2<1$}
Using $z=e^{ik}$, Eq.\ (\ref{al2}) is
\begin{eqnarray}
a_l &=& \frac{i\sqrt{ 1-\gamma^2}}{2\pi} \oint_{|z|=1}^{} \frac{dz}{z} z^{l} \frac{ \left(z-\alpha e^{i\psi}\right)^{1/2} \left(z-\alpha^{-1}e^{i\psi}\right)^{1/2} } {z^{1/2}}  \frac{ \left(z-\alpha e^{-i\psi}\right)^{1/2} \left(z-\alpha^{-1}e^{-i\psi}\right)^{1/2} } {z^{1/2}} ,
\end{eqnarray}
with the branch cuts along $(0, \alpha e^{i\psi}) \cup (\alpha^{-1}e^{i\psi}, \infty)$ and $\alpha e^{i\psi}=(\Gamma+i\sqrt{1-\Gamma^2-\gamma^2} )/(1+\gamma)$. That is, the phases are defined in the region $(\psi, 2\pi+\psi)$ for the branch cuts in the upper half-plane, and $(-\psi, 2\pi-\psi)$ for the cuts in the lower half-plane. 

\noindent By deforming the integration contour, we find
\begin{eqnarray}
a_l&=& \Re \left[\frac{2i\sqrt{1-\gamma^2}}{\pi} \int_0^1 \frac{dt}{t^2}\left(\alpha e^{i\psi}t \right)^l (1-t)^{1/2}
\right. \nonumber \\ &&\left.
 (\alpha^{-1}-\alpha t)^{1/2}  \left(t e^{i\psi}-e^{-i\psi} \right)^{1/2} \left(t\alpha e^{i\psi}-\alpha^{-1} e^{-i\psi} \right)^{1/2} \right]
 \label{al0}
\end{eqnarray}
To estimate the leading behavior in $N$, we expand the above integral around $t=1$ to get
\begin{eqnarray}
a_N&\approx& \Re \left[\frac{2i\sqrt{1-\gamma^2}}{\pi} \alpha^N e^{i\psi N} \int_0^1 \frac{dt}{t^2}t^N (1-t)^{1/2} (\alpha^{-1}-\alpha )^{1/2}  \left( e^{i\psi}-e^{-i\psi} \right)^{1/2} \left(\alpha e^{i\psi}-\alpha^{-1} e^{-i\psi} \right)^{1/2} \right] .
\nonumber \\
&=& \frac{4\sqrt{2}}{\pi} \left(\frac{\gamma^2 (1-\Gamma^2)(1-\Gamma^2-\gamma^2) }{1-\gamma^2} \right)^{1/4}\alpha^N  \cos(\psi N+\psi_0/2) \int_0^1 dt  (1-t)^{1/2} t^{N-2}
\nonumber \\
&\approx&\frac{4\sqrt{2}}{\pi} \left(\frac{\gamma^2 (1-\Gamma^2)(1-\Gamma^2-\gamma^2) }{1-\gamma^2} \right)^{1/4}\frac{\alpha^N}{2N\sqrt{N} }  \cos(\psi N+\psi_0/2) .
\end{eqnarray}
Therefore the asymptotic of the gap for large enough $N$,
\begin{eqnarray}
\Delta  &\approx& \left| \frac{4\sqrt{2}}{\pi} \left(\frac{\gamma^2 (1-\Gamma^2)(1-\Gamma^2-\gamma^2) }{1-\gamma^2} \right)^{1/4}\frac{\alpha^N}{\sqrt{N} }  \cos(\psi N+\psi_0/2) \right| .
\end{eqnarray}

\subsection{Isotropic critical line $\gamma=0$}
To that end we can take the limit $\gamma \to 0$ in Eq.~(\ref{al0}), which is equivalent to setting $\alpha=1$ and $\psi = \phi = \arccos(\Gamma)$.
\begin{equation}
a_l = \Re\left[\frac{2 i}{\pi} \int_{0}^{1}\frac{dt}{t^2} (e^{i\phi} t)^l (1-t) (t e^{i\phi} - e^{-i \phi}) \right].
\end{equation}

\noindent This integral can be calculated and subsequently expanded for large $l$ as,
\begin{equation}
a_l =\Re~\left[\frac{4 i (\cos \phi - i l \sin \phi ) }{\pi (l^2 -1) l} e^{i \phi l} \right] \simeq -4 \frac{\sqrt{1-\Gamma^2}}{\pi l^2} \cos (l \phi) + 4 \frac{\Gamma}{\pi l^3} \sin (l \phi) + \mathcal{O}(l^{-4}).
\end{equation}
Now, in the leading order in $N$ the gap is
\begin{equation}
\label{gapxxapp1}
\Delta \simeq \left|- 8 \frac{\sqrt{1-\Gamma^2}}{\pi N} \sum_{s=1,3,5,\dots} \frac{\cos(s N \phi)}{s^2} \right|. 
\end{equation}

\noindent The sum in the above expression gives the oscillating term periodic in $N \phi$ with the period $2 \pi$. Calculating the sum, we obtain
\begin{equation}
\Delta \simeq \left|\frac{\pi \sqrt{1-\Gamma^2}}{N}  \left( -\frac{2}{\pi} \arcsin(\cos (N\phi))\right) \right|,
\end{equation}
which can be also expressed using modulus as in Eq.\ \eqref{eq:isotropic_asymptotic}.

To complete the analysis we consider the next term in the expansion in powers of $1/N$. Extending Eq.~(\ref{gapxxapp1}) we get
\begin{equation}
\Delta \simeq \left| - 8 \frac{\sqrt{1-\Gamma^2}}{\pi N} \sum_{s=1,3,5,\dots} \frac{\cos(s N \phi)}{s^2}+ 8 \frac{\Gamma}{\pi N^2} \sum_{s=1,3,5,\dots} \frac{\sin(s N \phi)}{s^3} \right|. 
\end{equation}
The second term is also periodic in $N \phi$ with the period $2 \pi$, but the zeros are shifted by $\pi/2$.  It is piecewise quadratic in $\phi N$. The second term may also become more important when approaching the critical point at $\Gamma=1$, as suggested by the prefactors. Performing the sums, and writing the result in a compact form we obtain
\begin{eqnarray}
\Delta &\approx&\left| \frac{\pi \sqrt{1-\Gamma^2}}{N}  \left( -\frac{2}{\pi} \arcsin(\cos (N\phi))\right)  \right.
\nonumber\\ 
&& \left.+ \frac{\pi \Gamma}{4 N^2}  \left( \frac{2}{\pi} \arccos(\cos (N \phi)) \left(2-\frac{2}{\pi}\arccos(\cos (N \phi)) ~ \mathrm{Sign} (\sin (N \phi)) \right )\right) \right|.
\end{eqnarray}
For instance, this shows that the expansion in Eq.\ \eqref{eq:isotropic_asymptotic} does not capture the {\it exact} position of zeros of the energy gap.

\section{Relation between the gap and correlation function in the finite system}

We consider the ZZ correlation function of the one-dimensional quantum $XY$ model in a finite system in the ground state. It has been calculated in Ref.  \cite{mccoy},
\begin{eqnarray}
G_{z}^R(N , \Gamma , \gamma ) &\equiv& \langle \sigma_z^i \sigma_z^{i+R}  \rangle - \langle \sigma_z^i \rangle \langle \sigma_z^{i+R} \rangle
\nonumber \\
  &=&   \frac{1}{N^2} \sum_{k}  \left[ - \sin{\theta_{k}}  \sin( Rk)  + \cos {\theta_{k}} \cos(Rk)  \right] 
\nonumber \\
&&\times \sum_{k^{}}  \left[  \sin{\theta_{k^{}}}  \sin( Rk^{})  + \cos {\theta_{k^{}}} \cos(Rk^{})  \right] \label{G} ,
\end{eqnarray}
where
\begin{eqnarray}
k &=&
\begin{cases}
\displaystyle{ k_1 =0 , \pm \frac{2}{N} \pi ,\cdots ,  \pm \frac{N-2}{N} \pi , \pi} & (\text{$N_c$ odd}) 
\\
\displaystyle{k_2 = \pm \frac{1}{N} \pi , \pm \frac{3}{N} \pi , \cdots ,  \pm \frac{N-1}{N} \pi}  &  (\text{$N_c$ even}) 
\end{cases} \label{kk}
\end{eqnarray}
and
\begin{eqnarray}
\cos \theta_k &=& \frac{\cos k +\Gamma} { \sqrt[] { (\cos k+\Gamma)^2 + (\gamma \sin k)^2} } \label{costhetak}
\\
\sin \theta_k &=& \frac{\gamma \sin k} { \sqrt[] { (\cos k+\Gamma)^2 + (\gamma \sin k)^2} } .
\end{eqnarray}

Let us first consider the region $\Gamma^2+\gamma^2<1$.
We now show that the correlation function $G_{z}^N(2N , \Gamma , \gamma )$ is written explicitly in terms of the energy gap if $\Gamma^2+\gamma^2<1$ and the Fermion number is odd.
In this case, because $\sin (Nk_5)=0$ in Eq.\ (\ref{G}), the correlation function simplifies to
\begin{eqnarray}
G_{z}^N(2N , \Gamma , \gamma ) &=&   \frac{1}{4N^2} \left( \sum_{k_7} \cos {\theta_{k_7}} \cos(Nk_7) \right)^2
\nonumber \\
&=&\frac{1}{4N^2} \left( \sum_{k_1}  \cos {\theta_{k_1}} -\sum_{k_2}  \cos {\theta_{k_2}}  \right)^2,
\end{eqnarray}
where
\begin{eqnarray}
k_7 &=&0 , \pm \frac{1}{N} \pi , \pm \frac{2}{N} \pi ,\cdots ,  \pm \frac{N-1}{N} \pi , \pi
\\
k_1 &=&0 , \pm \frac{2}{N} \pi ,\cdots ,  \pm \frac{N-2}{N} \pi , \pi
\\
k_2 &=& \pm \frac{1}{N} \pi , \pm \frac{3}{N} \pi , \cdots ,  \pm \frac{N-1}{N} \pi  .
\end{eqnarray}
Furthermore, according to Eqs.\ (\ref{epsilon}) and (\ref{costhetak}), we find $\cos {\theta_k} =\partial \epsilon(k)/\partial \Gamma$,
and the correlation function is seen to have the following expression,
\begin{eqnarray}
G_{z}^N(2N , \Gamma , \gamma ) &=& \frac{1}{4N^2} \left( \frac{\partial}{\partial \Gamma} \left( \sum_{k_1} \epsilon(k_1) - \sum_{k_2} \epsilon(k_2) \right)  \right)^2 .
\end{eqnarray}
In the region $0 \le \Gamma \le 1$, from Eq.\ (\ref{eq:Delta2}), the gap is
\begin{equation}
\Delta(N , \Gamma , \gamma) = \left| - \sum_{k_1} \epsilon(k_1) + \sum_{k_2} \epsilon(k_2) \right| .
\end{equation}
We therefore have
\begin{eqnarray}
G_{z}^N(2N , \Gamma ,\gamma ) =\left( \frac{1}{2N}  \frac{\partial}{\partial \Gamma} \Delta(N , \Gamma , \gamma)  \right)^2 .
\end{eqnarray}

It is impossible to find a similar relation for an even number of Fermions because $\sin(Rk)$ in Eq.\ (\ref{G}) does not vanish.
Likewise, in the region $\Gamma^2 + \gamma^2>1 $,  because an even number of Fermions always gives the ground state, we are not able to find a similar relation.


\begin{thebibliography}{99}
\bibitem{nishimori-ortiz} H. Nishimori and G. Ortiz, {\em Elements of Phase Transitions and Critical Phenomena}, (Oxford, 2011).
\bibitem{santos} R. R. dos Santos and R. B. Stinchcombe, J. Phys. A: Math. Gen. {\bf 14}, 2741 (1981).
\bibitem{osterloh} A. Osterloh, Luigi Amico, G. Falci, and Rosario Fazio, Nature {\bf 416}, 608 (2002).
\bibitem{privman} V. Privman and M. E. Fisher, J. Stat. Phys. {\bf 33}, 385 (1983).
\bibitem{hatano} N. Hatano, Y. Nishiyama and M. Suzuki, J. Phys. A: Math. Gen. {\bf 27}, 6077 (1994).
\bibitem{kadowaki} T. Kadowaki and H. Nishimori, Phys. Rev. E {\bf 58}, 5355 (1998).
\bibitem{kadowaki2} T. Kadowaki, Thesis, Tokyo Institute of Technology (1998);
arXiv:quant-ph/0205020.
\bibitem{finilla} A. B. Finilla, M.A. Gomez, C. Sebenik, and D. J. Doll, Chem. Phys. Lett. {\bf 219}, 343 (1994).
\bibitem{morita} S. Morita and H. Nishimori, J. Math. Phys. {\bf 49}, 125210 (2008).
\bibitem{das} A. Das and B. K. Chakrabarti, Rev. Mod. Phys. {\bf 80}, 1061 (2008).
\bibitem{santoro} G. E. Santoro and E. Tosatti, J. Phys. A: Math. Gen. {\bf39}, R292 (2006).
\bibitem{bapst} V. Bapst, L. Foini, F. Krzakala, G. Semerjian, and F. Zamponi, Phys. Rep. {\bf523}, 127 (2013).
\bibitem{farhi} E. Farhi, J. Goldstone, S. Gutmann, J. Lapan, and D. Preda, Science {\bf 292}, 474 (2001).
\bibitem{seki} Y. Seki and H. Nishimori, Phys. Rev. E {\bf85}, 051112 (2012).
\bibitem{seoane} B. Seoane and H. Nishimori, J. Phys. A: Math. Theor. {\bf 45}, 435301 (2012).
\bibitem{bapst2} V. Bapst and G. Semerjian, J. Stat. Mech. 2012, P06007 (2012).
\bibitem{jorg} T. J\"{o}rg, F. Krzakala, J. Kurchan, A. C. Maggs, and P. Pujos, EPL {\bf 89}, 40004 (2010).
\bibitem{young1} A. P. Young, S. Knysh, and V. N. Smelyanskiy, Phys. Rev. Lett. {\bf 104},
020502 (2010).
\bibitem{young2} A. P. Young, S. Knysh, and V. N. Smelyanskiy, Comp. Phys. Commun. {\bf 182},
27 (2011).
\bibitem{cabrera} G. G. Cabrera and R. Jullien, Phy. Rev. Lett. {\bf57}, 393 (1986).
\bibitem{cabrera2} G. G. Cabrera and R. Jullien, Phys. Rev. B {\bf35}, 7062 (1987).
\bibitem{laumann} C. R. Laumann, R. Moessner, A. Scardicchio, and S. L. Sondhi, Phys. Rev. Lett. {\bf 109}, 030502 (2012).
\bibitem{tsuda} J. Tsuda, Y. Yamanaka, H. Nishimori, J. Phys. Soc. Jpn. {\bf 82}, 114004 (2013).
\bibitem{Henkel}  M. Henkel, J. Phys. A {\bf20}, 995 (1987).
\bibitem{sachdev} S. Sachdev, Quantum Phase Transitions (Cambridge : Cambridge University Press), (2011).
\bibitem{Damski} B. Damski and M. M. Rams, J. Phys. A: Math. Theor. {\bf 47}, 025303 (2014).
\bibitem{lieb} E. Lieb, T. Schultz, and D. Mattis, Ann. Phys. {\bf 16}, 407 (1961).
\bibitem{katsura} S. Katsura, Phys. Rev. {\bf 127}, 1508 (1962).
\bibitem{pfeuty} P. Pfeuty, Ann. Phys. {\bf 57}, 79 (1970).
\bibitem{mccoy} E. Barouch and B. M. McCoy, Phys. Rev. A {\bf 3}, 786 (1971).
\bibitem{kurmann} J. Kurmann, H. Thomas, and G. M\"{u}ller, Physica A {\bf 112}, 235 (1982).
\bibitem{hoeger} C. Hoeger, G. von Gehlen and V. Rittenberg, J. Phys. A: Math. Gen. {\bf 18}, 1813 (1985).
\bibitem{kenzelmann} M. Kenzelmann, R.Coldea, D. A. Tennant, D. Visser, M. Hofmann, P. Smeibidl, and Z. Tylczynski,
Phys. Rev. B {\bf 65}, 144432 (2002).
\bibitem{antonella} A. De Pasquale and P. Facchi, Phys. Rev. A {\bf 80}, 032102 (2009).
\bibitem{CV2010} M. Campostrini and E. Vicari, Phys. Rev. A {\bf 81}, 063614 (2010).
\bibitem{CNPV2014} M. Campostrini, J. Nespolo, A. Pelissetto, and E. Vicari,
Phys. Rev. Lett. {\bf 113}, 070402 (2014).
\bibitem{CPV2015} M. Campostrini, A. Pelissetto, and E. Vicari, Phys. Rev. E {\bf 91}, 042123 (2015).
\bibitem{mccoy2} B. M. McCoy, Phys. Rev. {\bf 173}, 531 (1968).
\bibitem{Zauner2015} V. Zauner, D. Draxler, L. Vanderstraeten, M. Degroote, J. Haegeman, M. M. Rams., V. Stojevic, N. Schuch and F.Verstraete, New J. Phys.  {\bf 17} 053002 (2015).
\end{thebibliography}
\end{document}